\documentclass[
prl,reprint,
,showpacs,twocolumn,floatfix,nofootinbib,superscriptaddress,
amsmath,amssymb,amsfonts,
nofootinbib
]{revtex4-1} 

\usepackage{bm}
\usepackage{slashed}
\usepackage{graphicx}

\usepackage{amsmath,amssymb,amsfonts,graphicx}
\usepackage{dsfont, mathrsfs}
\usepackage{bm}
\usepackage{slashed}
\usepackage{subfigure}
\usepackage[svgnames]{xcolor}
\usepackage{placeins}
\usepackage{array}
\usepackage{natbib,hyperref}  
\usepackage{url}

\makeatletter
\newcommand{\thickhline}{%
    \noalign {\ifnum 0=`}\fi \hrule height 1pt
    \futurelet \reserved@a \@xhline
}
\newcolumntype{"}{@{\hskip\tabcolsep\vrule width 1pt\hskip\tabcolsep}}
\makeatother

\usepackage{placeins}

\usepackage{xspace}

\newcolumntype{L}[1]{>{\raggedright\let\newline\\\arraybackslash\hspace{0pt}}m{#1}}
\newcolumntype{C}[1]{>{\centering\let\newline\\\arraybackslash\hspace{0pt}}m{#1}}
\newcolumntype{R}[1]{>{\raggedleft\let\newline\\\arraybackslash\hspace{0pt}}m{#1}}
\newcommand\SlimSec[1]{\textbf{\textit{#1}}---}
\newcommand{\Eq}[1]{Eq.~(\ref{#1})}
\newcommand{\al}[1]{\begin{align} #1 \end{align}}
\newcommand{\non}{\nonumber}
\newcommand{\vect}[1]{\boldsymbol{#1}}

\def\vf{\varphi}
\def\fh{\varphi_h}

\def\fS{\varphi_S}
\def\fR{\varphi_R}
\def\fRB{\varphi_{\bar{R}}}

\def \fk{\varphi_{k}}

\def\fKR{\varphi_{KR}}
\def\fKRB{\varphi_{\bar{K} \bar{R}}}
\def \fq{\varphi_{q}}
\def \fqR{\varphi_{qR}}
\def \fqRB{\varphi_{q\bar{R}}}

\def\kT{\vect{k}_T}
\def\kBT{\vect{\bar{k}}_T}
\def\RT{\vect{R}_T}

\def\RBT{\vect{\bar{R}}_T}
\def\qT{\vect{q}_T}

\def\pT{\vect{p}_T}

\newcommand{\ImL}{1\columnwidth}
\newcommand{\GapCapt}{\vspace{-8pt}}

\begin{document}

\title{Accessing Quark Helicity through Dihadron Studies} 

\preprint{ADP-17-42/T1048}

\author{Hrayr~H.~Matevosyan}
\thanks{ORCID: http://orcid.org/0000-0002-4074-7411}
\affiliation{ARC Centre of Excellence for Particle Physics at the Tera-scale,\\ 
and CSSM, Department of Physics, \\
The University of Adelaide, Adelaide SA 5005, Australia
\\http://www.physics.adelaide.edu.au/cssm
}

\author{Aram~Kotzinian}
\thanks{ORCID: http://orcid.org/0000-0001-8326-3284}
\affiliation{Yerevan Physics Institute,
2 Alikhanyan Brothers St.,
375036 Yerevan, Armenia
}
\affiliation{INFN, Sezione di Torino, 10125 Torino, Italy
}

\author{Anthony~W.~Thomas}
\thanks{ORCID: http://orcid.org/0000-0003-0026-499X}
\affiliation{ARC Centre of Excellence for Particle Physics at the Tera-scale,\\     
and CSSM, Department of Physics, \\
The University of Adelaide, Adelaide SA 5005, Australia
\\http://www.physics.adelaide.edu.au/cssm
}

\begin{abstract}
We present a new proposal to study the helicity-dependent dihadron fragmentation functions (DiFF), which describe the correlations of the longitudinal polarization of a fragmenting quark with the transverse momenta of the produced hadron pair. Recent experimental searches for this DiFF via azimuthal asymmetries in back-to-back hadron pair production in $e^+e^-$  annihilation by the {\tt BELLE} Collaboration did not yield a signal. Here we propose a new way to access this DiFF in $e^+e^-$ annihilation, motivated by the recently recalculated cross section of this reaction, which explains why there was in fact no signal for the {\tt BELLE} Collaboration to see. In this new approach  the azimuthal asymmetry is weighted by the virtual photon's transverse momentum square multiplying sine and cosine functions of difference of azimuthal angles of relative and total momentum for each pair. The integration over the virtual photon's transverse momentum has the effect of separating the convolution between the helicity-dependent DiFFs in the quark and antiquark jets and results in a nonzero collinear expression containing Fourier moments of helicity-dependent DiFFs. A second new measurement is also proposed for two-hadron production in semi-inclusive deep inelastic scattering, where the asymmetry is weighted in a similar way for a single pair. This results in a collinear factorized form of the asymmetry, which includes the quark helicity parton distribution function and the same helicity-dependent DiFF as in $e^+e^-$ production and will allow us to check the universality of this DiFF.
\end{abstract}



\date{\today}                                           

\maketitle

\label{SEC_INTRO}
 
Understanding quark hadronization remains one the most challenging problems in modern hadronic physics. In hard inclusive reactions, this process is encoded by the so-called fragmentation functions (FF) for the detection of a single semi-inclusive final state hadron and by the so-called dihadron fragmentation functions (DiFF) for hadron pair production. In the case of the hadronization of an unpolarized quark where the observed hadrons are unpolarized, these fragmentation functions can be interpreted as probability densities for observing the given type of hadron. When the initial quark is polarized, the modulations of the azimuthal distributions of the final state hadrons can act as polarimeters.  This is a consequence of correlations between the polarization of the quark and the transverse momenta of the produced hadrons. In fact, such correlations have been experimentally observed between the transverse polarization of the quark and the transverse momenta of the hadrons, both for one hadron (Collins effect) and two hadrons (interference DiFF) in semi-inclusive production. On the other hand, such correlations have not yet been observed for a longitudinally polarized quark, which is only possible when at least two unpolarized hadrons are detected. In this Letter we address this issue by proposing two new experimental measurements of such correlations, the first in $e^+e^-$ annihilation and the second  in semi-inclusive deep inelastic scattering (SIDIS).

 The measurements of two-hadron azimuthal asymmetries in the semi-inclusive process have been recently used to access the quark transversity parton distribution function (PDF) inside the nucleon~\cite{Bacchetta:2011ip,Bacchetta:2012ty,Radici:2015mwa}, using a combined analysis of two-hadron production in SIDIS and back-to-back two-hadron pair creation in $e^+e^-$ annihilation. A key role here is played by  the so-called interference DiFF (IFF), that describes a correlation between the transverse spin polarization of a fragmenting quark with the relative transverse momentum of the produced hadron pair. Similarly, the helicity-dependent  DiFF, $G_1^\perp$,  describes a correlation between the longitudinal polarization of a fragmenting quark and the transverse momenta of the pair of hadrons. Its importance lies to a considerable extent in its relationship to the phenomenon of longitudinal jet handedness, which was predicted 25 years ago~\cite{Efremov:1992pe} but has not yet been observed. It is also of interest because it has no analog in single unpolarized hadron production. 
 
  The first proposed azimuthal asymmetry for measuring $G_1^\perp$, in back-to-back  two-hadron pair creation in $e^+e^-$, was made over ten years ago in~\cite{Boer:2003ya}. A subsequent experimental search for this asymmetry by the {\tt BELLE} Collaboration collaboration did not yield a signal~\cite{Abdesselam:2015nxn, Vossen:2015znm}, while their previous measurements for the IFF signal were sizable~\cite{Vossen:2011fk}. The model calculations of the specific form of the integrated $G_1^\perp$ entering this asymmetry was recently performed in~\cite{Matevosyan:2017alv}, producing a result that is smaller than that for IFF calculated within the same model, but still non-negligible.
  
  Recently, motivated by the findings in~\cite{Matevosyan:2017uls}, we rederived the cross section expressions for dihadron production in  $e^+e^-$ annihilation~\cite{Matevosyan:2018icf}, and found a number of disagreements with the previous calculations. The two most important conclusions were the resolution of the apparent inconsistencies between the definitions of IFF entering the two mentioned processes and the realization that the originally proposed azimuthal asymmetry for  determining $G_1^\perp$ in $e^+e^-$ annihilation should vanish.
  
In this Letter, we propose a new measurement to access $G_1^\perp$ in $e^+e^-$ annihilation, based on the new expression for the cross section. An experimental search using this method would be very important to gain any knowledge on $G_1^\perp$.  Such information is vital in our understanding of the hadronization process, and together with the measurements of IFF provide a critical test for the models. Moreover, we also propose a  new measurement in SIDIS, which will give access to  $G_1^\perp$ multiplied by the collinear nucleon helicity PDF, which itself is well determined from a number of inclusive and semi-inclusive measurements. Thus, the measurements of the asymmetries both in $e^+e^-$  and SIDIS will allow us to test the universality of $G_1^\perp$ entering  these two processes. We note that the universality of the unintegrated DiFFs entering in both SIDIS and $e^+e^-$ reactions has not yet been explicitly proven when including the gauge link in the quark fragmentation correlator, though the universality of the DiFFs integrated over the total transverse momenta of the hadron pair was demonstrated in Ref.~\cite{Bacchetta:2003vn}.
 
\label{SEC_EE_XSEC}
\begin{figure}[t]
\centering 
\includegraphics[width=\ImL]{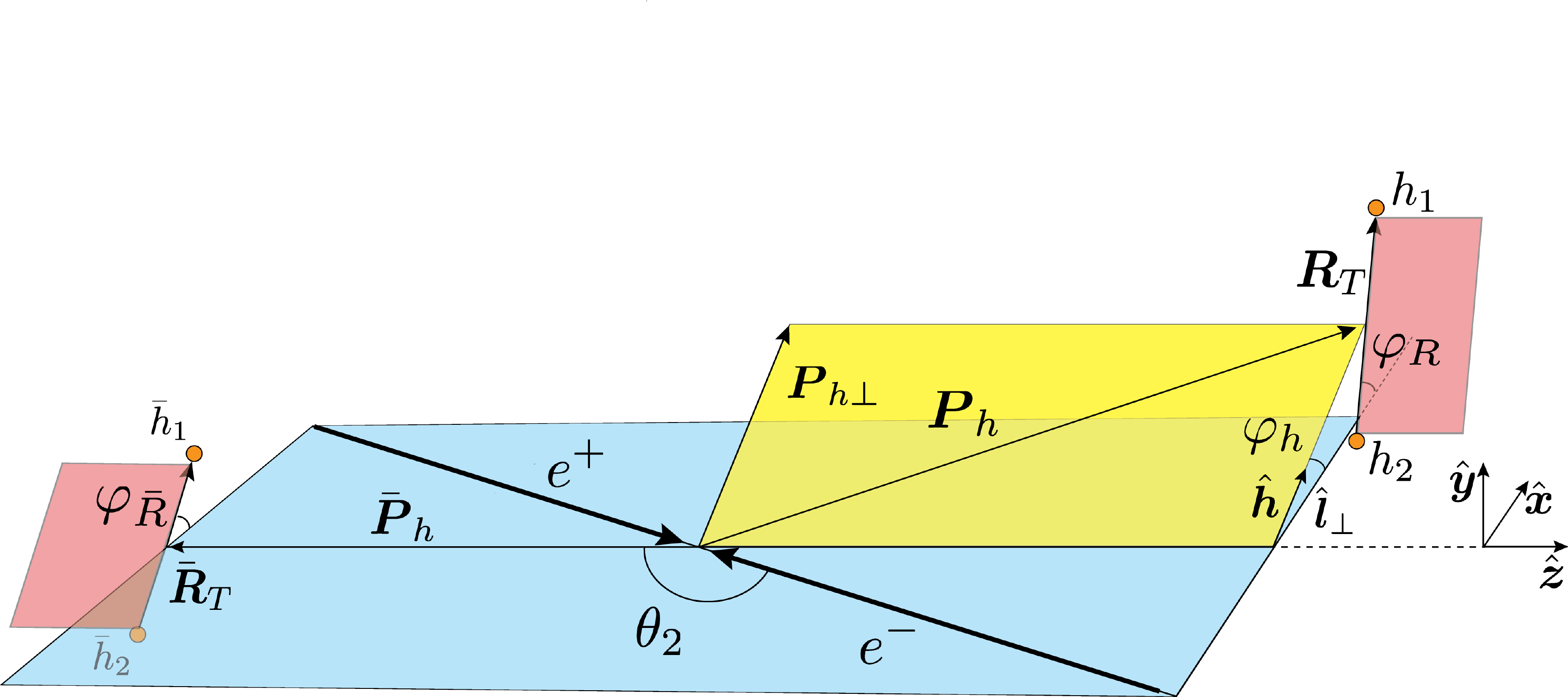}
\GapCapt
\caption{The kinematics of two back-to-back dihadron pair creation in $e^+e^-$ annihilation.}%
\label{PLOT_EE_KINEMATICS}
\vspace{-0.1cm}
\end{figure}

\SlimSec{The $e^+e^-$  asymmetry.}\-The process we consider is $e^+e^-\to h_1 h_2 + \bar{h}_1 \bar{h}_2 + X$, where the initial electron and positron are unpolarized, and the two-hadron pairs $h_1 h_2$  and $\bar{h}_1 \bar{h}_2$ are emitted back to back. The electron and the positron carry momenta $l$ and $l'$, respectively. The final state hadrons $h_{1}, h_{2}$ are assigned momenta $P_1,P_2$ and masses $M_1,M_2$, while the second hadron pair $\bar{h}_1,\bar{h}_2$ are assigned momenta  $\bar{P}_1,\bar{P}_2$ and masses $\bar{M}_1,\bar{M}_2$. The total and the relative momenta of the hadron pairs are defined as $ P \equiv P_h = P_1 + P_2$, $ R = \frac{1}{2}( P_1 - P_2)$, while their invariant mass is defined as $M_h^2 = P_h^2$.
 In the leading order approximation for the hard scattering part of this process, the $e^+e^-$ pair annihilates into a virtual photon with momentum $q = l + l'$. In turn, this decays into a quark and antiquark $a,\bar{a}$ carrying momenta $k$ and $\bar{k}$.  The hard scale $Q$ in this work is defined by $Q^2=q^2$ and is assumed to be much smaller than the $Z$ boson mass. In the following we will ignore all the contributions of order $1/Q$. The quark and antiquark hadronize and produce two back-to-back jets of particles, each containing one of the hadron pairs considered here. The other kinematical variable entering the cross section is
{
  $y = {P_h \cdot l}/{P_h \cdot q} \approx {l^-}/{q^-}$,
}
where the light-cone components of a four-vector $a$ are defined as $a = (a^+,a^-,\vect{a}_T)$, and  $a^\pm = \frac{1}{\sqrt{2}}(a^0 \pm a^3)$. The cross section also depends on the the light-cone momentum fractions of the produced hadrons $z_i = {P_i^+}/{k^+}$ and their total and relative combinations
{
$z = z_1  + z_2$
and $ \xi = {z_1}/{z}$.
}
The coordinate system used in the analysis is defined in the center of mass frame of colliding $e^+e^-$ by taking the $\hat{z}$ opposite to the three-momentum $\vect{\bar{P}}_h$, and the components of the vectors perpendicular to $\hat{z}$ are denoted with a subscript $_\perp$,  see Fig.~\ref{PLOT_EE_KINEMATICS}. In this frame $\vect{q}_\perp=0$. It is also useful to define a reference frame where the total momenta of both hadron pairs are collinear~\cite{Boer:2003ya}, and the components of three-vectors perpendicular to them is denoted by $_T$. In this frame the virtual photon has a transverse momentum component $\vect{q}_T = -\vect{{P}}_{h\perp} /{z}$, while the difference between the $_T$ and $_\perp$ components of the observed hadron momenta is  of the order $1/Q$ and will be neglected here. Note, that the azimuthal angle of $\qT$ and $\vect{P}_h$ are related as $\fq= \vf_h+ \pi$.

 The cross section expression for this process at leading twist was originally derived in~\cite{Boer:2003ya}, using the "leading hadron approximation" $P_h \cdot \bar{P}_h \sim Q^2$. It contains convolutions of DiFFs encoding the hadronization of the quark and the antiquark. The new derivation, in~\cite{Matevosyan:2018icf}, corrected several errors in the previous result. The relevant part of the cross section here contains the unpolarized and helicity-dependent DiFFs, and the azimuthal dependence is determined by
\al
{
\label{EQ_XSEC}
&\frac{d \sigma_{U,L}^{e^+e^- \to (h_1 h_2) (\bar{h}_1 \bar{h}_2) X }}{d^2 \qT \ d\fR\  d\fRB
\ d^7 V 
} 
 = \sum_{a,\bar{a}} e_a^2 \frac{3 \alpha^2}{\pi Q^2} z^2 \bar{z}^2  A(y)
\\\non
&\times
\Big\{
\mathcal{F}\Big[ D_1^a \bar{D}_1^{\bar{a}}\Big]
- \mathcal{F}\Big[\frac{(\RT\times \kT)_3}{M_h^2} \frac{(\RBT \times \kBT)_3}{\bar{M}_h^2}  G_1^{\perp a} \bar{G}_1^{\perp \bar{a}}  \Big]
 \Big\},
}
where $\fR, \fRB$ are the azimuthal angles of $\RT$ and $\RBT$. The symbol $d^7V\equiv dz d\xi d R_T d \bar{z} d\bar{\xi}  d \bar{R}_T dy$ denotes the remaining phase space element. The subscript $3$ denotes the $z$ component of the vector.
The electromagnetic coupling constant is denoted  as $\alpha$, while the charges of the quarks are $e_a$, and 
{
$A(y) = \frac{1}{2} - y + y^2$.
}
The convolution $\mathcal{F}$ is defined as
\al
{
\label{EQ_F_CONVOL}
\mathcal{F}[w D^a \bar{D}^{\bar{a}} ]
 =  \int d^2 \kT &d^2 \kBT 
 \  \delta^2(\vect{k}_T + \vect{\bar{k}}_T - \vect{q}_T)
\\ \non
 & \times 
  w( \kT, \kBT, \RT, \RBT)\ D^a \ D^{\bar{a}}.
}
The fully unintegrated DiFFs entering these expressions depend only on the relative azimuthal angles between $\kT$, $\RT$ and  $\kBT$, $\RBT$, respectively: $D^a(z, \xi, \kT^2, \RT^2, \kT \cdot \RT  )$, $D^{\bar{a}}(\bar{z}, \bar{\xi}, \kBT^2, \RBT^2, \kBT \cdot \RBT  )$.

In order to gain information about the DiFFs entering the cross section, it is helpful to use their decomposition into Fourier cosine series with respect to the relative azimuthal angle $\fKR \equiv \fk - \fR$
\al
{
\label{EQ_D_FOURIER}
D^a(z, \xi, \kT^2, \RT^2,& \cos(\fKR)) 
\\ \non
  =& \frac{1}{\pi} \sum_{n=0}^\infty
 \frac{\cos(n \cdot \fKR)}{1+\delta^{0,n}} \ D^{a,[n]}(z, \xi, \kT^2, \RT^2),
}

The unweighted integrated cross section over $\qT, \fR, \fRB, \xi, \bar{\xi}$ contains only the unpolarized DiFFs
\al
{
\label{EQ_XSEC_INT_UPOL}
 \langle 1 \rangle 
 &=  \int d \sigma_{U,L}^{e^+e^- \to (h_1 h_2) (\bar{h}_1 \bar{h}_2) X }\  \times 1
 \\ \non
& =   \frac{3 \alpha^2}{\pi Q^2}  A(y)
 \sum_{a,\bar{a}} e_a^2   D_1^a(z, M_h^2) \bar{D}_1^{\bar{a}}(\bar{z}, \bar{M}_h^2),
}
where the integrated zeroth Fourier moment of the unpolarized DiFF is defined as
\al
{
D_1^{a}(z, M_h^2)  =  z^2 \int d^2 \kT \int  d \xi
\ D_1^{a,[0]}(z, \xi, \kT^2, \RT^2) \, .
}
Here, $\RT^2$ has been replaced by the invariant mass square $M_h^2$ of the the hadron pair, per convention. 
It is easy to see that the previously proposed measurement of an asymmetry containing $G_1^\perp$ in~\cite{Boer:2003ya} simply vanishes
\al
{
\langle \cos(2(\fR - \fRB)) \rangle =0.
}
In fact, it is easy to demonstrate that $\langle f(\fR, \fRB)\rangle=0$ for an arbitrary $f$ depending only on $R$ and $\bar R$.

We note that we can extract information about the term containing $G_1^\perp$ in the cross section in \Eq{EQ_XSEC}, by evaluating a moment that would contain a weight of $\sin(\fKR) \sin(\fKRB)$, multiplied by any cosine Fourier harmonics of the same angles. In the experimental analysis, this can be achieved by employing various combinations of the measurable azimuthal angles of $\fq=\vf_h+ \pi$ and $\fR$, $\fRB$. For example, $\sin(m \fqR ) \sin(n \fqRB)$, where $\fqR \equiv \fq-\fR$ and $\fqRB \equiv \fq-\fRB$, should yield a nonvanishing result involving convolutions of various Fourier cosine moments of $G_1^{\perp a}$ and $G_1^{ \perp \bar{a}}$ for any $m,n>0$. Such weighted integrals of the cross section should in principle also  contain Fourier moments of the unpolarized DiFFs, but no contributions form transverse polarization dependent DiFFs. The convolution in \Eq{EQ_F_CONVOL}, containing the $\delta^2(\vect{k}_T + \vect{\bar{k}}_T - \vect{q}_T)$ function from transverse momentum conservation, can be factorized into a product of weighted Fourier moments of the helicity-dependent DiFFs by introducing an additional weighting factor $q_T^2$. The contribution of $D_1$ moments can be canceled out by forming linear combinations of different asymmetries. For example, for the case $m=1$, $n=1$
\al
{
\label{EQ_PHI1_AV}
&\left\langle \frac{q_T^2
\big(3 \sin(\fqR) \sin(\fqRB) 
+ \cos(\fqR) \cos(\fqRB) \big)
}{M_h \bar{M}_h  }
 \right \rangle
\\ \non
&
= \frac{12 \alpha^2 A(y) }{\pi Q^2} 
 \sum_{a,\bar{a}}  e_a^2  
 \Big( G_1^{\perp a,[0]}  - G_1^{\perp a,[2]} \Big) \Big( \bar{G}_1^{\perp \bar{a}, [0]}  - {G}_1^{\perp \bar{a}, [2]} \Big),
}
where the dimensionless integrated $n$th moments are
\al
{
\label{EQ_G_MOMS}
G_1^{\perp a,[n]}(z, M_h^2) \equiv  &\ z^2  \int d^2 \kT  \int  d \xi 
\\ \non
&\times
 \left(\frac{\kT^2}{2 M_h^2}\right)  \frac{|\RT|}{M_h} \ G_1^{\perp a,[n]}(z, \xi, \kT^2, \RT^2).
}
The combination of the zeroth and the second Fourier cosine moments arises from the trigonometric relation $\sin^2 (\fKR) = (1-\cos(2\fKR))/2$, where  both terms couple to the corresponding Fourier cosine moment in the  decomposition~(\ref{EQ_D_FOURIER}).
For convenience, we  define the integrated helicity-dependent DiFF as
\al
{
\label{EQ_G1_INT}
G_1^{\perp a}(z, M_h^2) \equiv  G_1^{\perp a,[0]}(z, M_h^2)  - G_1^{\perp a,[2]}(z, M_h^2).
}
Note, that this definition differs from that in~\cite{Boer:2003ya}. Our recent model calculations of Fourier cosine moments of $G_1^\perp$ in~\cite{Matevosyan:2017alv} suggest a sizable analyzing power for such a combination, although in the model calculations we used $k_T R_T$ weighting when defining the Fourier cosine moments of $G_1^\perp$ instead of $k_T^2 R_T$ used in \Eq{EQ_G_MOMS}.

The corresponding azimuthal asymmetry, which is the ratio of the weighted moment in \Eq{EQ_PHI1_AV} to the unweighted one in \Eq{EQ_XSEC_INT_UPOL}, can be expressed as
\al
{
\label{EQ_ee_SSA}
 A_{e^+e^-}^{\Rightarrow}(z ,& \bar{z}, M_h^2, \bar{M}_h^2)
 = 4
\frac{ \sum_{a,\bar{a}} 
 G_1^{\perp a}(z, M_h^2) \ G_1^{\perp \bar{a}}(\bar{z}, \bar{M}_{h}^2)
 }
 { \sum_{a,\bar{a}} 
 D_1^a(z, M_h^2) \ D_1^{\bar{a}} (\bar{z}, \bar{M}_{h}^2) 
 }.
}

The idea of using transverse-momentum weighting to break up the momentum convolutions in single hadron azimuthal asymmetries was first employed a number of years ago in~\cite{Kotzinian:1995cz,Kotzinian:1997wt} and later in~\cite{Boer:1997nt}. The first experimental results motivated by that work have only just been released by the {\tt COMPASS} Collaboration~\cite{Bradamante:2017yia,Matousek:2017xpc}. In recent years, an improved method of Bessel-weighted asymmetries has been proposed in~\cite{Boer:2011xd} and tested for viability using Monte Carlo simulations in~\cite{Aghasyan:2014zma}. Though our proposed asymmetry can be also expressed in terms of Bessel-weighted functions, we leave that for future work.

\SlimSec{The azimuthal asymmetries in SIDIS.}\-We next consider the SIDIS process with two observed final state hadrons $l + N \to l' + h_1h_2 +X$. We use the expression for the fully unintegrated cross section of this process, derived in~\cite{Bacchetta:2002ux}, to suggest another weighted asymmetry involving $G_1^\perp$.

Here we use the standard kinematics of SIDIS~\cite{Bacchetta:2002ux}, where the initial and final state leptons are assigned momenta $l$ and $l'$, the initial nucleon $N$ has mass $M$, momentum $P$, and polarization $\vect{S}$. The final state hadrons are again assigned momenta $P_1$, $P_2$ and masses $M_1$, $M_2$. The single-photon exchange approximation  is used, where the momentum of the intermediate virtual photon is $q= l-l'$ and the hard scale is defined by $Q^2 = -q^2$. The  $\gamma^*N$ center of mass coordinate system is chosen, where $\hat{z}$ axis is taken along the three-momentum of the virtual photon $\vect{q}$, and the $\hat{x}$ axis along the the transverse momentum of leptons. In the parton picture, a quark with momentum $p$ absorbs the virtual photon, acquiring momentum $k = p + q$. This quark then hadronizes, producing the two observed hadrons in a jet of particles. The relevant variables are the light-cone momentum fraction of the initial quark $x = p^+/P^+$, its transverse momentum $\pT$, as well as the transverse momentum of the final quark $\kT$. The light-cone momentum fractions of the final state hadrons are defined with respect to $k^-$, that is $z_{1,2} = P_{1,2}^-/k^-$. The relative and the total transverse momenta and light-cone momentum fractions are defined in the same manner as for $e^+e^-$ annihilation. Using a Lorentz transform, it can be easily shown that $\qT = - \vect{P}_{h\perp}/z$. Finally, the momentum fraction $y$  is defined here as $y = (P\cdot q)/(P \cdot l) \approx l^-/q^-$.

The cross section for this process can be decomposed into various terms according to the polarization of the incident lepton beam and the target nucleon. The two cases of interest here are $\sigma_{UU}$ and $\sigma_{UL}$, describing the unpolarized and target longitudinal polarization dependent parts of the cross section, respectively. Here we only show the  explicit dependence of these two terms on the relevant azimuthal angles  
\al
{
\label{EQ_SIDIS_XSEC_UU}
\frac{d \sigma_{UU} }{ d^2 \vect{P}_{h\perp} d\fR  \ d^6V'}
=  \sum_a \frac{\alpha^2 e_a^2}{\pi y Q^2} A'(y) \
 \mathcal{G} \Bigg[ f_{1}^a\ D_1^{ a}
 \Bigg],
}
and
\al
{
\label{EQ_SIDIS_XSEC_OL}
&\frac{d \sigma_{UL} }{ d^2 \vect{P}_{h\perp} d\fR \ d^6V'}
\\\non
&\hspace{1cm}
= -  S_L \sum_a \frac{\alpha^2 e_a^2}{\pi y Q^2}  A'(y)
 \mathcal{G} \Bigg[\frac{(\RT \times \kT)_3}{M_h^2} g_{1L}^a  G_1^{\perp a} \Bigg],
}
where $d^6V' \equiv dz d\xi dM_h^2 dx dy d\fS$, $\fS$ is the azimuthal angle of the initial  nucleon's transverse polarization $\vect{S}_T$, $A'(y) = 1- y + y^2/2$, $S_L$ is the longitudinal polarization of the nucleon, and the SIDIS convolution is defined as
\al
{
\label{EQ_G_CONVOL}
\mathcal{G}[&w f^q D^{q} ] \equiv \int d^2 \pT  \int d^2 \kT
   \delta^2\Big(\kT -\pT+ \frac{\vect{P}_{h\perp}}{z} \Big)
\\ \non   
&\times    w( \pT, \kT, \RT)
f^q(x, p_T) D^{q}(z, \xi, k_T^2, R_T^2, \kT \cdot \RT  ).
}

The unweighted integral of the cross section over $ \vect{P}_{h\perp}, \fR,\xi, \fS$ yields a product of the unpolarized PDF and the DiFF
\al
{
& \left\langle 1 \right\rangle 
 = \int d \sigma_{UU} \times 1
=  \sum_a \frac{2 \alpha^2 e_a^2}{y\ Q^2} A'(y) f_1^a(x) z^2 D_1^a (z, M_h^2).
}

To access $G_1^\perp$, we again need to weight the cross section with a trigonometric factor containing only the first Fourier sine mode $\sin(\fR-\fk)$ and an arbitrary Fourier cosine mode $\cos(m(\fR-\fk))$, $m\geq 0$. The simplest modulation containing the observable angles, that would result in such a combination would be $\sin(\fh - \fR)$. Here again, by weighting this modulation by a factor of $P_{h\perp}$, we can break up the transverse momentum  convolution in~(\ref{EQ_G_CONVOL}) into a product of two independent terms
\al
{
 \left\langle \frac{P_{h\perp} \sin(\fh - \fR)}{M_h} \right\rangle 
 &= \int d \sigma_{UL}    \frac{P_{h\perp} \sin(\fh - \fR)}{M_h}
\\ \non
&\hspace{-3.2cm}=   S_L  A'(y)\sum_a \frac{2 \alpha^2 e_a^2}{y\ Q^2} 
\  g_{1}^a(x) \ z\ G_1^{\perp a}(z, M_h^2),
}
where
\al
{
 g_{1}^a(x) = \int d^2 \pT \ g_{1L}^a(x, p_T^2),
}
is the nucleon collinear helicity PDF, while the combination of the Fourier cosine moments of the helicity-dependent DiFF,  $G_1^{\perp a}(z, M_h^2)$, is exactly that appearing in the $e^+e^-$ annihilation asymmetry in \Eq{EQ_G1_INT}.

Thus, the proposed azimuthal asymmetry can be expressed as 
\al
{
\label{EQ_SIDIS_SSA}
 A^{\Rightarrow}_{SIDIS}(x, z , M_h^2)
 =&S_L \frac{\sum_{a}  g_{1}^a(x)
 \ z\ G_1^{\perp a}(z, M_h^2) }
 { \sum_{a} 
f_1^a(x)\ D_1^a(z, M_h^2)  }.
}
%
 
\SlimSec{Conclusions.}\-In this Letter we have proposed two new measurements which will permit us to access the helicity-dependent DiFF, both in back-to-back two-hadron pair production in $e^+e^-$ annihilation and in forward two-hadron production in SIDIS. In both cases, the proposed asymmetries measure a combination of the zeroth and second cosine Fourier moments of the integrated $G_1^\perp$ and we demonstrated a prescription for accessing the higher Fourier cosine moments of $G_1^\perp$ by way of using different azimuthal modulations.  In SIDIS this is multiplied by the nucleon helicity PDF. This presents a very promising opportunity for testing the universality of this DiFF between SIDIS and $e^+e^-$ processes, since the helicity PDF has been experimentally determined to a high precision. This is in contrast to the IFF, which is coupled with the poorly known transversity PDF. Indeed the universality of IFF has been used to extract transversity via a combined analysis of SIDIS and $e^+e^-$ measurements. The proposed $e^+e^-$measurements can be accomplished by reanalyzing the the {\tt BELLE} Collaboration data used in previous searches of $G_1^\perp$ as well as through new measurements in the upcoming the {\tt BELLE} Collaboration~II experiment. The complimentary SIDIS measurements could be done at Jefferson Lab 12 and the proposed Electron-Ion Collider. Once measured, $G_1^\perp$ can serve as a polarimeter to determine quark longitudinal polarization in a number of other process, such as hadron pair production in polarized $pp$ scattering.

The work of H.H.M. and A.W.T. was supported by the Australian Research Council through the ARC Centre of Excellence for Particle Physics at the Terascale (CE110001104), and by the ARC  Discovery Project No. DP151103101, as well as by the University of Adelaide. A.K. was supported by the A.I. Alikhanyan National Science Laboratory (YerPhI) Foundation, Yerevan, Armenia.

\bibliographystyle{apsrev4-1}
\bibliography{fragment}

\end{document}